\documentclass{pasj00}

\begin{document}
\SetRunningHead{M.Matsuoka et al.}{The MAXI mission on the ISS}
\Received{2009/03/09}
\Accepted{2009/05/28}

\title{
The MAXI Mission on the ISS: Science and Instruments \\
for Monitoring All Sky X-Ray Images
}

\author{Masaru \textsc{Matsuoka}, \altaffilmark{1}
        Kazuyoshi \textsc{Kawasaki}, \altaffilmark{1}
        Shiro \textsc{Ueno}, \altaffilmark{1}
        Hiroshi \textsc{Tomida}, \altaffilmark{1}
        \\
        Mitsuhiro \textsc{Kohama}, \altaffilmark{1,2}
        Motoko \textsc{Suzuki}, \altaffilmark{1}
        Yasuki \textsc{Adachi}, \altaffilmark{1}
        Masaki \textsc{Ishikawa}, \altaffilmark{1}
        \\
        Tatehiro \textsc{Mihara}, \altaffilmark{2}
        Mutsumi \textsc{Sugizaki}, \altaffilmark{2}
        Naoki \textsc{Isobe}, \altaffilmark{2}
        Yujin \textsc{Nakagawa}, \altaffilmark{2}
        \\
        Hiroshi \textsc{Tsunemi}, \altaffilmark{3}
        Emi \textsc{Miyata}, \altaffilmark{3}
        Nobuyuki \textsc{Kawai}, \altaffilmark{4}
        Jun \textsc{Kataoka}, \altaffilmark{4}%
        \thanks{Present Address: Research Institute for Science and Engineering, Waseda University, 3-4-1, Okubo, Shinjuku, Tokyo, 169-8555}
        \\
        Mikio \textsc{Morii}, \altaffilmark{4}
        Atsumasa \textsc{Yoshida}, \altaffilmark{5}
        Hitoshi \textsc{Negoro}, \altaffilmark{6}
        Motoki \textsc{Nakajima}, \altaffilmark{6}
        \\
        Yoshihiro \textsc{Ueda}, \altaffilmark{7}
        Hirotaka \textsc{Chujo}, \altaffilmark{2}
        Kazutaka \textsc{Yamaoka}, \altaffilmark{5}
        Osamu \textsc{Yamazaki}, \altaffilmark{5}
        \\
        Satoshi \textsc{Nakahira}, \altaffilmark{5} 
        Tetsuya \textsc{You}, \altaffilmark{5}
        Ryoji \textsc{Ishiwata}, \altaffilmark{6}
        Sho \textsc{Miyoshi}, \altaffilmark{6}
        \\
        Satoshi \textsc{Eguchi}, \altaffilmark{7}
        Kazuo \textsc{Hiroi}, \altaffilmark{7}
        Haruyoshi \textsc{Katayama}, \altaffilmark{8}
        and
        Ken \textsc{Ebisawa}, \altaffilmark{9}
        }
\altaffiltext{1}{ISS Science Project Office, ISAS, JAXA, 2-1-1 Sengen, 
Tsukuba, Ibaraki 305-8505}
\email{(MM) matsuoka.masaru@jaxa.jp}
\altaffiltext{2}{Cosmic Radiation Laboratory, RIKEN, 2-1 Hirosawa, 
Wako, Saitama 351--198}
\altaffiltext{3}{Department of Earth and Space Science, Osaka University ,
1-1 Machikaneyama, 
\\
Toyonaka, Osaka 560-0043}
\altaffiltext{4}{Department of Physics, Tokyo Institute of Technology, 
2-12-1 Ookayama, 
\\
Meguro-ku, Tokyo 152-8551}
\altaffiltext{5}{Department of Physics and Mathematics, Aoyama Gakuin University, 
\\
5-10-1 Fuchinobe, Sagamihara, Kanagawa 229-8558}
\altaffiltext{6}{Department of Physics, Nihon University, 
1-8-14, Kanda-Surugadai, 
\\
Chiyoda-ku, Tokyo 101-8308}
\altaffiltext{7}{Department of Astronomy, Kyoto University, Oiwake-cho, 
\\
Sakyo-ku, Kyoto 606-8502}
\altaffiltext{8}{Earth Observation Research Center, JAXA, 2-1-1 Sengen, 
\\
Tsukuba, Ibaraki 305-8505}
\altaffiltext{9}{ISAS, JAXA, 3-1-1 Yoshinodai, Sagamihara, Kanagawa 229-8510}



%

\KeyWords{ASM, All Sky Monitor, X-ray nova, AGN, GRB, X-ray transient, 
X-ray source catalogue} 

\maketitle

\begin{abstract}
   The Monitor of All Sky X-ray Image (MAXI) mission is the first 
astronomical payload to be installed on the Japanese Experiment Module 
- Exposed Facility (JEM-EF or Kibo-EF) on the International Space 
Station (ISS).  It is scheduled for launch in the middle of 2009 to 
monitor all-sky X-ray objects on every ISS orbit.  
It will be more powerful than any previous X-ray All Sky  Monitor  (ASM)
 payloads, being able to monitor hundreds of Active Galactic Nuclei 
(AGNs).  A realistic simulation under optimal observation conditions
 suggests that MAXI will provide all sky images of X-ray sources of 
$\sim$20 mCrab ( $\sim$7$\times$10$^{-10}$ ergs 
cm$^{-2}$ sec$^{-1}$ in the 
energy band of 2-30 keV) from observation on one ISS orbit 
(90 min), $\sim$4.5 mCrab for one day, and $\sim$ 2 mCrab for one week. 
The final detectability of MAXI could be  $\sim$0.2 mCrab for 
two years, which is comparable to the source confusion limit  of the 
MAXI field of view (FOV).  The MAXI objectives are  (1) to  alert the
community to X-ray novae and 
transient X-ray sources, (2) to monitor long-term variabilities of X-ray sources, 
(3) to stimulate multi-wavelength observations of variable objects, (4) to create 
unbiased X-ray source catalogues, and (5) to observe diffuse cosmic X-ray emissions, 
especially with better energy resolution for soft X-rays down to 0.5 keV.

   MAXI has  two types of X-ray slit cameras with
wide FOVs and two kinds of  X-ray detectors consisting of 
gas proportional counters covering the energy range of 2 to 30 keV and X-ray CCDs covering 
the energy range of 
0.5 to 12 keV.  Both cameras scan all-sky X-ray images twice due to
two different directional cameras  every 90 minutes synchronized 
with the ISS orbit. The data are sent through the downlink  between 
the ISS and a ground station via data-relay satellites.  MAXI  will thus enable 
us to report  X-ray novae or transients to astronomers worldwide within 
a few minutes.  The ground-based nova-alert system rapidly reports these events  
with 0.1 to 0.2 degree position accuracy   to astronomers worldwide for further follow-up 
observations.  Measurements
of each source are combined on the ground.  As a result, we  
will be able to detect even a weak source such as an AGN.  MAXI is capable of creating 
source catalogues for specific periods to investigate the variabilities of X-ray sources.

\end{abstract}

\section{Introduction}

The All Sky Monitor (ASM) for X-ray observations has a long history 
(Holt and Priedhorsky 1987).
The ASM on Ariel 5 (a British satellite) was the first dedicated 
pioneer payload to observe X-ray novae and transients (Holt 
1976).  The Ariel 5 ASM, which has two sets of one-dimensional scanning pinhole 
cameras,  discovered several novae and transient X-ray objects. 
Thereafter, the terms "X-ray nova" and "X-ray transient" have become 
well-known in X-ray astronomy.  X-ray instruments with a wide field of 
view (FOV) as well as ASM can not only detect X-ray novae and transients, 
but also monitor long-term X-ray variabilities of X-ray sources 
(Priedhorsky and Holt 1987).

Although Vela A \& B satellites with a wide FOV discovered 
the first gamma-ray burst (GRB) before the Ariel-ASM (Klebesadel et al. 
1973), those detectors were not much suitable for monitoring GRBs
as well as X-ray 
novae and transients. Subsequently, dedicated GRB monitors confirmed 
that a considerable number 
of mysterious GRBs are produced in the universe. GRB monitors with a wide FOV 
have advanced greatly since a wide field  camera on Beppo-SAX 
discovered a GRB afterglow (Costa et al. 1997).  Since then, special GRB 
satellites such as HETE-2 (Ricker et al. 2002; Shirasaki et al. 
2004) and Swift (Gehrels et al. 2004) have been realized.  

The  ASM, however,  advanced gradually as a supplemental 
payload.  Ariel-ASM with pin-hole cameras  operated successfully 
for seven years with the main payload which made spectrum observations 
of bright X-ray sources (Holt 1976).   In 1987 the ASM on the Japanese 
satellite Ginga (Tsunemi et al. 1989) succeeded  the Ariel-ASM.  
Ginga-ASM was able to observe the spectra of X-ray novae as it 
scanned the sky from 60 degrees through 360 degrees once every day with 
multi-slat collimators. Ginga-ASM greatly  advanced the science of black 
hole binaries with the discovery of nova-like black 
hole binaries (Tsunemi et al. 1989; Kitamoto et al. 1992; Kitamoto et al. 
2000). Ginga-ASM 
was operated successfully for 4.5 years, and some of transients 
discovered by ASM were observed in detail by the main large area counters 
on Ginga (Turner et al. 1989).

Since 1996, RXTE-ASM (a NASA satellite) has monitored known 
X-ray sources in addition to X-ray novae and transients (Levine et al. 
1996).  Systematic long-term data of Galactic variable sources provide 
quasi-periodic properties of the  accretion disc (Zdziarski et al. 
2007a; 2007b).  RXTE-ASM has provided much useful data for X-ray variable 
sources for 13 years, but the detection limit is around 10 mCrab.  
Therefore, the main targets of RXTE-ASM were the Galactic X-ray sources 
of which monitoring was suitable for performing detailed observations
of these targets with an RXTE prime instrument, a large-area proportional counter array 
 (Remillard and McClintock 2006).

Now we have long term light curves for bright X-ray sources from 
the above three ASMs.  Some sources revealed  periodic or quasi-periodic
component of long time scale with combined analysis over the last 
30 years 
(Paul et al. 2000).  Future ASMs taking over the three ASMs can  
continue to investigate further long-term behaviour of X-ray sources.  
Although the above three ASMs have been in operation, they cannot 
 adequately  monitor AGNs because of their low sensitivity.  Thus far, 
Galactic X-ray novae and transients data  have been accumulated 
although the information is not completely clear.  No long term
 variability of AGN has not been observed with sufficient data. Therefore,
 the most important role  for future ASMs is to better the detection 
limit  for AGN monitoring.  Considering this historical situation, 
MAXI was proposed  as  a payload of ISS (Matsuoka et al. 1997).  
It will be
one of the ASMs responsible for AGN monitoring.  A slit hole camera 
with a large detection area meets this requirement.  Since MAXI has 
a composite structure of slit holes and slat collimators without a 
mirror system, the angular 
resolution is not  good (i.e., the FWHM of slat collimators is 
 1.5 degrees), but the localization accuracy can be 
0.1 degree for the sources of enough statistic (e.g., bright sources
and sources with enough counts accumulated for optimum time).  
A mirror-type ASM, 
such as the Lobster-eye ASM, is promising for the future although
 its energy band  is limited to soft X-rays
  (Priedhorsky et al. 1996).

The MAXI to be attached to the Japanese Experiment Module (JEM; Kibo) on the 
ISS is ready 
to be launched.  In this paper we will present an overview of MAXI 
and the new scientific contributions expected of it.  
The first part of this paper will discuss the astronomical science expected of
MAXI. The second part  will  describe the MAXI instrumentations,
 and observational simulation.

\section{Key Science}

The MAXI mission enables the  investigation of ASMs and surveying. 
 MAXI will alert astronomers of GRBs, X-ray novae, and 
flare-up increases of X-ray sources 
if they occur. Long-term data of X-ray sources will enable us to 
determine a special time scale of variability, e.g., long-term periodic
or quasi-periodic motions of X-ray sources.  MAXI can promote multi-wavelength 
observation  in collaboration with other space and ground 
observatories such as X-ray, infrared, and optical satellites, 
radio, and optical ground observatories. MAXI systematic observations 
of the variable activity of black hole binaries and AGNs 
are used to investigate how and 
where they produce their variable activities. 

MAXI provides unbiased X-ray source catalogues over  
all the sky.  Monthly or biannual 
X-ray catalogues could contribute to the long-term study of variable 
behavior of AGN for the 
first time.  A catalogue accumulated for two years could provide 
all-sky AGNs corresponding to a moderately deep survey to the 
entire sky. 
This unbiased AGN catalogue will far surpass the investigation
of  the 
distribution and evolution of AGNs for the HEAO-1 A2 catalogue which 
has been utilized for long time (Piccinotti et al. 1982). MAXI is also 
able to make an all sky X-ray map with soft X-rays and medium 
energy X-rays.  The soft X-ray map provides  line features such 
as Oxygen X-ray lines, which are useful in researching geo-coronal 
recombination lines (Fujimoto et al. 2007) as well as the evolution of hot gas 
in the Galaxy (McCammon and Sanders 1990; Tanaka and Bleeker 1977). 

Current Swift (Gehrels et al. 2009; Burrow 2009) and INTEGRAL (Ubertini 
et al. 2009) satellites with wide FOVs of hard X-ray detectors have 
provided X-ray source catalogues including AGNs in addition to 
a considerable number of transients.  Since these results are 
complementary to those of MAXI,  it could be promising to science
 that these three ASM missions will operate simultaneously.
Furthermore, the recently developed gamma-ray large-area 
space telescope, the Fermi Gamma-Ray Space
Telescope (Abdo et al. 2008; Thompson et al. 2009), is a gamma-ray ASM
 with an energy band complementary to that of MAXI.  All-sky images with
 both ASMs may reveal new information.

\subsection{X-ray Novae and GRBs Alert}

 More than 90\% of the black hole candidates are X-ray transients or 
novae (McClintock and Remillard 2007).  In the last 20 years, 29 
have been discovered, mostly with  RXTE-ASM and partly  with Ginga-ASM, while 
about 30\% of them have been serendipitously discovered by
pointing observatories (Negoro 2009).  Thus far, observed 
X-ray novae are located 
near the Galactic center and near the solar system; i.e., their 
distances are within 
 about 6 kpc.  Distant X-ray novae appear weak,
 but MAXI could 
 detect these weak sources, since it can observe X-ray novae from 
distant regions about three times as far.  Thus, it is expected 
that the nova discovery
 rate may increase by one order of magnitude.  

 X-ray light curves with energy spectra of X-ray novae have provided 
 instability and phenomena 
 deviating from the standard accretion disc model (Tanaka and Shibazaki 
 1996). Data of both low and high luminosities have resulted in 
the creation of a new 
 theory concerning accretion discs (Mineshige et al. 1994), but these 
 samples are not enough.  We also expect that MAXI could observe 
 X-rays from classical novae.  Multi-wavelength observations of classical 
 novae could provide useful information about the bursting 
mechanism of classical novae.  Although classical novae are faint
and soft in X-ray band,  the durations are of months to years
(Mukai, Orio \& Della Valle 2008).  Thus MAXI could detect some of 
them.

 Swift has discovered many GRBs and has consequently enabled us to make rapid 
 follow-up observations of them.  Nevertheless, the energy band of 
 the Swift detector is  15 to 200 keV (Gehrels et al. 
 2002).  MAXI can cover the soft X-ray band of GRBs, and this 
information is important to the understanding of  X-ray rich 
GRBs  and X-ray flashes.  It is expected that the population of the 
X-ray rich GRB is comparable to that of the GRB with an ordinary 
energy band (Sakamoto et al.  2005).  The instant 
FOV of MAXI is not  large, MAXI will be able to detect
3.5 prompt emissions and 2.5 X-ray afterglows of GRB per year
 (Suzuki et al. 2009). 

  MAXI is also able to observe  transient  X-ray binary pulsars 
with high orbital eccentricity  and transient  low mass X-ray binaries with neutron stars.  
Furthermore, Anomalous X-ray Pulsars (AXPs) are still  enigmatic
objects that appear as Soft Gamma Repeaters (SGRs) with enormous flux
during a short time (Nakagawa et al. 2009; Morii 2009).  
Considering the recent
Swift discovery of a new SGR (Barthelmy et al. 2008; Enoto et al. 2009), we could 
even expect MAXI to
detect additional SGRs with weaker flux.  Thus, MAXI may reveal new frontier of
AXPs (i.e., magnetars) by detecting numerous new magnetar candidates 
(Nakagawa et al. 2009).
Recently, special interest concerning long type I X-ray bursts has 
arisen with rare events of  carbon-fueled super
bursts and helium-fueled intermediately long bursts. These bursts are 
promising in the investigation of the  deeper neutron star 
envelope (Keek and in't
 Zand 2008; Keek et al 2008).  MAXI  may discover such rare 
occurrences of X-ray bursts.

\subsection{Long-Term Variability of X-Ray Sources}

  Most X-ray sources with compact objects are variable, due to 
the instability of the accretion disc and other  reasons.
Periodic or quasi-periodic long-term variability is sometimes 
created from interaction between the accretion disc and the binary system.  
 If there 
 were a third object in the X-ray sources, we could expect some other 
 periodicity in addition to binary period (Zdziarski 2007a).  The 
third body might be useful for inspecting the accretion disc.
However, no one has 
 discovered such a triple-body system in the X-ray sources.  A composite 
 time scale of variability would help generate this knowledge and/or
 may create new knowledge.
An unusual X-ray transient from the Galactic center region
(Smith et al. 1998) is progressing in a new class of recurrent 
and fast X-ray transient sources (i.e., SFXTs).  
These transients sometimes 
occur in high-mass X-ray binaries associated with super-giant companions.   It is promising for further
investigation  that MAXI in addition to INTEGRAL and Swift monitors
these objects with a wide FOV(Ebisawa 2009; Ubertini et al. 2009) .
  Generally, a time scale of AGNs variability is longer than that of 
 X-ray binaries.  The time scale of AGNs is useful in understanding a 
 complicated structure around the AGNs.  If some AGN had formed a 
 super-massive binary black hole system, we could expect some 
 periodic variation over a year's time (Hayasaki et al. 2008).

\subsection{Multi-Wavelength Observations of Variable Objects}

  The knowledge of bursts, transients, and variable objects has 
 progressed considerably with multi-wavelength observations.  
 Research of GRB afterglows has advanced greatly as a result of rapid 
 follow-up observations in X-ray, optical, and radio bands 
(Gehrels et al. 2006), but there are still just a few samples of
short GRBs and X-ray rich GRBs.  
 Coordinate observations of Blazars in radio, infra-red, optical, 
 X-ray, and ultra-high-energy (TeV) gamma-rays have confirmed a 
Synchrotron Self-Compton (SSC) model for highly variable periods 
(Kubo et al. 1998; Kataoka et al. 1999; Takahashi et al. 2000).  
 However, the emission mechanism of other AGNs as well as Blazars 
 has not yet been completely  understood because of complicated 
 correlations among multi-wavelength observations to
help investigate this mechanism (Maoz et al. 2002). 
  TeV gamma-ray observations by \'Cerenkov telescopes have progressed 
 remarkably since TeV gamma-rays from some Blazars were 
 detected (Petry et al. 1996; Aharonian et al. 1997).  GeV gamma-ray 
observations by Fermi-GLAST are now available as an ASM (Thompson et al. 2009).  
MAXI  can promote further simultaneous multi-wavelength observations 
of Galactic active objects as well as AGNs with  gamma-rays (GeV \& TeV) to 
radio bands (Fender 2009; Madejsky et al. 2009).

\subsection{Unbiased X-Ray Source Catalogues}

  Nominal beam size (i.e., angular resolution by FWHM) of MAXI is 
 1.5$\times$1.5 deg$^{2}$.  It is estimated that a confusion limit 
 of X-ray sources is 5$\times$10$^{-12}$ erg cm$^{-2}$ sec$^{-1}$ from
 recent 
 Log N - Log S plot of X-ray sources (Ueda et al. 2003); i.e., 0.2 
 mCrab in the energy band of 2 to 20 keV.  Thus,  we can set
 a detection limit of 0.2 mCrab as an ideal goal, although the 
 period of its achievement depends on intrinsic background and 
 systematic error.  To estimate observation time we  
 conducted a realistic MAXI observational simulation.  MAXI  
 achieved a detection limit of 0.2 mCrab with a two-year observation 
(Ueda et al. 2009).  
 Other simulations suggest that it is possible to detect 20 mCrab 
 for one orbit (90 min.), 4.5 mCrab for one day, and 2 mCrab for one 
 week (Hiroi et al. 2009; Sugizaki et al. 2009). Some regions in 
the sky are covered by a bright region  from the Sun and are 
sometimes affected by the South Atlantic Anomaly (SAA).  Therefore,  
  detection sensitivities are estimated under the best 
 condition.  The detectability of MAXI is not uniform in 
 the entire sky, but slightly depends on the direction.  
  The MAXI simulation indicates that we can obtain 
 30-40 AGNs every week (Ueda et al. 2009).  This sample is 
comparable to the number from 
 HEAO 1-A2  (Piccinotti et al. 1982).  We are also able to estimate 
 about 1000 AGNs with two-year observations.  MAXI  observes those 
AGNs in a harder energy band than observed by ROSAT.  Thus, MAXI 
 could give an unbiased population ratio of Type I and Type II 
 AGNs.   Furthermore, if weekly or monthly catalogues are created, 
 we can discover the intensity variability for a considerable number of 
 AGNs. It is possible to detect a flare-up of Blazars and then  
 follow their light curves.
   The time sequence of X-ray unbiased catalogues is very useful 
in researching 
  the evolution of AGNs as well as their long-term variability. 
Here the next comment is noted.   Although much deeper surveys
of AGNs in the 2-10 keV band with ASCA (e.g., Ueda et al. 1999) and 
Chandra/XMM-Newton (Brandt \& Hasinger 2005 and references therein) 
have presented Log N -- Log S plots for researching the evolution
of AGNs, they are limited to a small portion of the  sky, and hence 
cannot constrain that of bright AGNs with small surface densities.

\subsection{Diffuse Cosmic X-Ray Emissions}

  The problem of the diffuse cosmic X-ray background (CXB)
 has a long history.  
A recent deep survey of X-ray sources in a medium energy band of 2 to 10 
 keV (Ueda et al. 2003) suggests that the CXB may be due to the 
superposition of AGNs, as proposed 
 theoretically (Morisawa et al. 1990).   It is still unknown  
 how some kinds of AGNs are distributed in the universe.  A global CXB 
 distribution is compared with that of optical AGNs and/or an infrared 
 map. Thus we can investigate the evolution or emission mechanism of the 
 all-sky AGN distribution. If we can obtain some difference of 
distribution in different energy bands, we can 
 investigate the distribution of Type I and Type II AGNs.

  The diffuse emission of soft X-rays less than 1 keV is attributed to 
 geo-coronal gas as well as hot bubbles from supernova remnants 
(Tanaka and Bleeker 1977; McCammon and Sanders 1990). 
 ROSAT obtained a precise all sky map with a broad energy band 
 of soft X-rays.  MAXI can also 
 obtain all-sky maps of soft X-rays but with better energy 
 resolution (e.g., resolving Oxygen K-line, and Neon 
 K-line).  Recent Suzaku observations revealed a 
 strong contribution of recombination K-lines of Oxygen and 
 Carbon in the geo-coronal region (Fujimoto et al. 2007).  MAXI is 
 able to observe the Oxygen  line with seasonal variation 
 and solar activity. Therefore, MAXI observations 
 make it possible to discriminate between the contribution of geo-corona  
 and that of supernova remnants. Thus, we could investigate geo-coronal 
 science as well as element-evolution of supernova remnants from 
 soft X-ray line observations. 

\section{MAXI Project}

  The large-scale Space Station (SS) project started  in 
1985 with the collaboration of the USA, Japan, Canada, and the 
European Space Agency (ESA).  Planning
of the JEM  was also initiated at that time.  In 
1995, Russia began to take part in the SS project. The project was 
subsequently reduced to its present scale and then renamed the 
International Space Station (ISS).  JEM planning and 
designing have continued, but actual construction and basic tests 
of the engineering model began at that time. JEM consists of a 
pressurized module for micro-gravity experiments and an exposed 
facility (EF). 
\begin{table*}
\begin{center}
\caption{Specification of MAXI slit cameras}\label{tab:camera}
\begin{tabular}{lcc}
\hline
   & GSC\footnotemark[$\dagger$]: Gas Slit Camera & SSC\footnotemark[$\dagger$]: Solid-state slit camera \\
\hline
\\
X-ray detector
		& {\parbox{55mm}{
		 12 pieces of one-dimensional PSPC;
		  Xe + CO$_2$ 1 \%}}
			& {\parbox{55mm}{
			 32 chips of X-ray CCD;
                   \\ 
			 1 square inch, 1024x1024 pixels}}
\\
X-ray energy range
		& 2--30 keV
			& 0.5--12 keV
\\
Total detection area
		& 5350 cm$^2$
			& 200 cm$^2$
\\
Energy resolution
		& 18 \% (5.9 keV)
			& $\leq$ 150 eV (5.9 keV)
\\
Field of view\footnotemark[$*$]  
		& 1.5 $\times$ 160 degrees
			& 1.5 $\times$ 90 degrees
\\
Slit area for camera unit
		& 20.1 cm$^2$
			& 1.35 cm$^2$
\\
Detector position resolution 
		& 1 mm 
			& 0.025 mm (pixel size)
\\
Localization accuracy 
		& 0.1 deg
			& 0.1 deg
\\
Absolute time resolution
		& 0.1 msec(minimum)
			& 5.8 sec(nominal)
\\
Weight\footnotemark[$\ddagger$] 
		& 160 kg
			& 11 kg
\\
\hline
\multicolumn{3} {@{}l@{}}{\hbox {\parbox{165mm}{\footnotesize
   Notes.  
\par\noindent
\footnotemark[$*$]  FWHM $\times$ Full-FOV.
\par\noindent
\footnotemark[$\ddagger$] MAXI total weight: 520 kg.
\par\noindent
\footnotemark[$\dagger$] GSC consists of 6 camera units, where each unit
  consists of two PCs.  SSC consists of two camera units, SSC-Z and SSC-H.
}}}
\end{tabular}
\end{center}
\end{table*}
\begin{figure}
\begin{center}
\FigureFile(85mm,){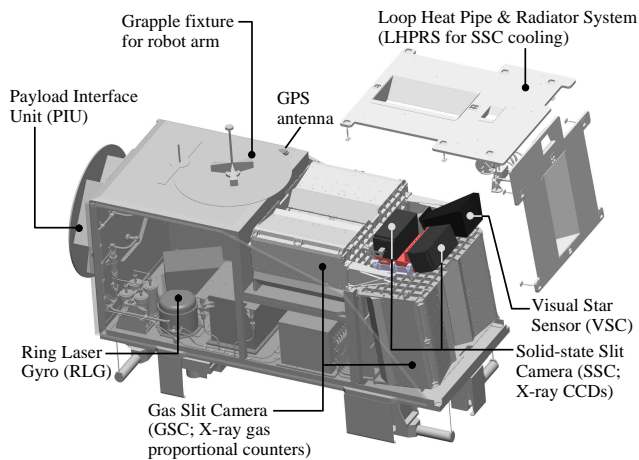}
\end{center}
\caption{Overview of MAXI; major subsystems are indicated.}\label{fig:overview}
\end{figure}

 ISS rotates synchronously in its orbit so that one side always 
points towards the center of the Earth and the opposite side views 
the sky.  Therefore, the sky side of JEM-EF surveys a great circle 
every ISS orbit.  Themes for JEM-EF include the space 
science payload for astrophysics and Earth observations, and space 
technology experiments such as robotics. However,  
JEM-EF does not provide a perfectly stable platform because of 
unknown factors of some attitude fluctuation. Payload size 
and weight are limited to the capacity of the
JEM-EF.  A payload can be suitable for 
survey observation, but not for pointing.  
Although a considerably wide FOV is available, 
the ISS structure and solar paddle partially block the view of JEM-EF. 
Since the ISS carries various experimental instruments and 
payloads, each instrument and payload has to accommodate many interfaces.  
Resources such as communication and power for each experiment or 
payload are also limited.  

Considering these problems MAXI was 
proposed and finally accepted in 1997 by the National Development
Space Agency of Japan (NASDA), now known as the Japan
Aerospace Exploration Agency (JAXA).  Thus, MAXI is the first 
astronomical payload for JEM-EF on the ISS. Although the launch was 
scheduled for 2003, the space shuttle, Columbia accident, and the ISS 
construction delay resulted in the postponement of the MAXI launch  
 by several years.   MAXI science instruments consist of two types of  
X-ray cameras, the Gas Slit Camera (GSC) and the Solid-state 
Slit Camera (SSC). These instruments 
and the support instruments on the MAXI payload are 
shown in Figure 1, and 
their characteristics are listed in Table 1.  The support instruments 
consist of a Visual Star Camera (VSC), a Ring Laser Gyroscope (RLG), a
Global Positioning System (GPS) and a Loop Heat Pipe and Radiation 
System (LHPRS).  The VSC  and the
RLG determine the directions of the GSC and the SSC as precisely 
as a few arc-seconds every second.  The GPS attaches the absolute 
time as precisely as 
0.1 msec to GSC photon data.  The LHPRS is used for heat transportation 
and heat radiation from thermo-electric coolers
(Peltier elements) to cool the CCD. 

The main role of the GSC is to perform as the X-ray ASM, which has the
best detectability than previous ASMs have  for a time scale
longer than hours.  All sky X-ray images obtained by GSC are  also
useful for new X-ray variable catalogues.  On the other hand the main 
mission of SSC is to make all sky X-ray maps for extended sources
with better energy resolution than ROSAT maps, although the 
detection area and live time for discrete sources are less by 
1/20 than those of GSC.  We also expect to detect the transients 
in soft X-ray band although the detectability of transients with 
SSC is also poorer by 1/20 than that of GSC.

 MAXI will be carried by the Space Shuttle, Endeavour, along with the JEM (or Kibo)- EF from Kennedy Space Center in the middle of  
2009.  MAXI will finally be mounted  on JEM-EF within two weeks 
of Endeavour launch.  At that time all of Kibo's  modules will 
have been installed on ISS.  After the basic arrangement of the 
structure and infrastructure of JEM, MAXI will conduct performance 
tests for about three months.  MAXI has a nominal lifetime of two 
years, but the expected goal is five or more years to achieve long-term 
monitoring.

 MAXI data will be down linked through the Low Rate Data Link (LRDL; 
MIL1553B), and the Medium Rate Data Link (MRDL; Ethernet).  MAXI is operated through the 
Operation Control System (OCS) at Tsukuba Space Center (TKSC), JAXA.  
MAXI data are not only processed and analyzed at TKSC, but they are 
also transferred to the Institute of Physical and Chemical Research
(RIKEN) MAXI data facility.  General users of MAXI can request the 
scientific data from RIKEN (a main port) as well as from JAXA (a sub-port) 
whenever they desire.  To search where X-ray 
novae, transients, or flaring phenomena appear suddenly in the sky 
we will conduct automatic data analysis at TKSC by using down 
linked data from 
LRDL.  If  such a source is discovered, we will report this event to 
astronomers and dedicated users worldwide via the Internet
from TKSC.  This nova alert system has been  developed  to achieve
automatic alert (Negoro et al. 2008). 
 The MAXI team will also maintain archival data at RIKEN using the data from 
LRDL and MRDL, such as all sky X-ray images, 
X-ray light curves, and spectra of dedicated X-ray sources. 
In principle all astronomical data of MAXI will be available
for public distribution (Kohama et al. 2009).


\section{X-ray Mission Instruments}
\begin{figure}
\begin{center}
\FigureFile(45mm,){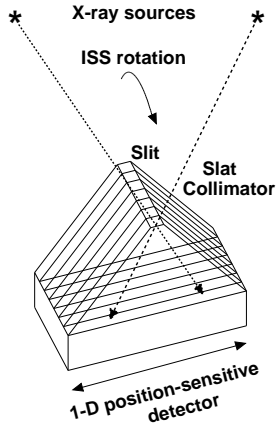}
\end{center}
\caption{A principle of MAXI Slit Camera; it consists of slit 
\& slat collimators and one-dimensional position sensitive 
X-ray detector.}\label{fig:camera}
\end{figure}

\begin{figure}
\begin{center}
\FigureFile(60mm,){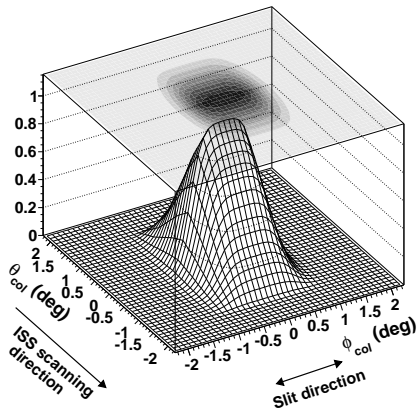}
\end{center}
\caption{Typical sample of PSF (Point Spread Function); a slat 
collimator has a triangular response (ISS scanning direction), 
while a slit collimator has a trapezoidal response (Slit direction corresponding to $\beta$ in Fig.6).}\label{fig:psf}
\end{figure}

 A specific object is observed by MAXI with a slit camera for a 
limited time with every ISS orbit. 
The X-ray detector of the camera is sensitive to a one 
dimensional image through the slit, where the wide FOV 
through the slit spans the sky perpendicular to the ISS moving 
direction as shown in Figure 2.  The scanning image of an 
 object is obtained 
with the triangular response of a slat collimator according to ISS 
movement. An intersection  of the slit image and the triangular response 
image corresponds to a source location in the sky as a Point Spread 
Function (PSF).  This is shown  in Figure 3. 

 Objects located along a great circle stay for 45 sec in the 
FOV of MAXI cameras, where the time of stay is the shortest in the 
MAXI normal direction (for a great circle). For  objects in the slanted 
 field from the great 
circle, they achieve slightly longer observation.  
Any target will come 
repeatedly in each field of the horizontal and zenithal cameras with 
every ISS orbital period. In this situation MAXI can intermittently 
monitor a short time scale variability for bright 
X-ray sources such as X-ray pulsars and low mass X-ray binaries.  
It can monitor their variability of weak sources on a time 
scale of 90 min or longer if we integrate the data.
MAXI has  two types of X-ray cameras: the GSC and the SSC which are
described in the following sub-section.

\subsection{Gas Slit Camera (GSC)}

\begin{figure}
\begin{center}
\FigureFile(85mm,){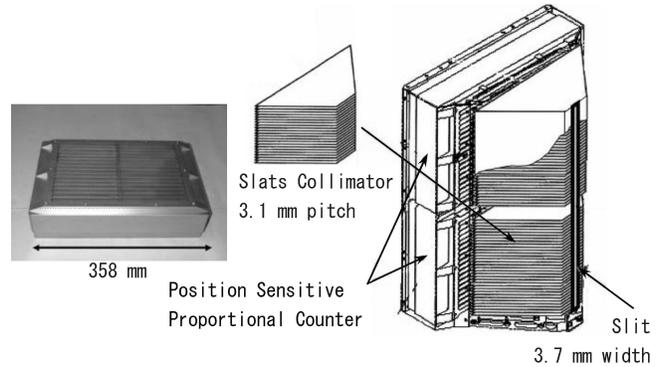}
\end{center}
\caption{ The GSC unit and a gas-proportional counter; the unit
 consists of two gas-proportional counters, slit and slat 
collimators.
}\label{fig:gscu}
\end{figure}

  GSC is the main X-ray camera and consists of six units of a 
 conventional slit camera as shown in Figure 1 and Table 1
 (Mihara et al. 2009).  
 The GSC unit consists of two one-dimensional proportional 
 counters (produced by Oxford Instruments Co. in Finland), and 
slit \& slat  collimators as shown in Figure 4.   Thus, twelve 
 proportional counters have a 5350 cm$^{2}$ detection area in total.  
Each  counter has   six cells of resistive carbon wires that are 
 guarded by a veto-detector region in the bottom and on both sides 
(Mihara et al. 2001), while the carbon wire divides the charge from 
the signal  into both terminals for one-dimensional determination.  
The X-ray  detection efficiency of the proportional counter (Xe 1.4 
atmospheres with 1 $\%$ CO$_2$) is 
plotted against  X-ray energy in Figure 5.  Observational energy
band of GSC is set to be 2-30 keV in standard operation mode 
where the detection efficiency for X-rays in this band
is above 10 $\%$ as seen in Figure 5.   The observation above Xe K-edge
of 34.6 keV  is possible as a special mode by adjusting amplifier
gain and high voltage.  Although a response function above Xe K-edge
 is complicated due to escape peak, the test for this function has
 been done using synchrotron X-ray beams.  The slat collimator response 
is 3.5  degrees in bottom-to-bottom, and a slit 
 image corresponds to 1.5 degrees covering a wide field between $-$ 40 
 degrees and $+$ 40 degrees.  The proportional counters detect 
 incident X-ray photons from vertical to $\pm$ 40 degree 
slant direction. 

Two FOVs of GSC-H and GSC-Z are placed in horizontal (forward) 
and zenithal directions to compensate when the sky is unobservable to
one FOV or the other 
(due to  high radiation background, such as SAA). 
The FOV of one camera is 80 degrees. Two camera units 
can cover 160 degrees, but the FOV of each central camera of GSC-H
and GSC-Z overlaps with each half of the FOV of both side cameras to
even the exposures as shown in Figure 6 which shows the product  
of effective area and dwell time according to one orbit scan. 
The exposure time per orbit depends on the direction of a star 
by a factor of  1/cos$\beta$, where $\beta$ is the angle of the star
slanted from  scanning  a great circle. 
Both edges of 10 
degrees in this figure are omitted, due to the shadow of the ISS structure, 
where the  scanning loss is 1.5$\%$. 
The forward FOV is tilted up by 6 degrees so as not to observe the Earth even 
when the ISS attitude changes. Thus, the FOV of GSC-H can scan the sky 
with ISS rotation without Earth occultation.  By making the observation time
longer than 90 min, we made the instrument simpler, 
and then optimized it to search in the AGN discovery space for the longer-term 
AGN variability ($>$ 1.5 hour).
\begin{figure}
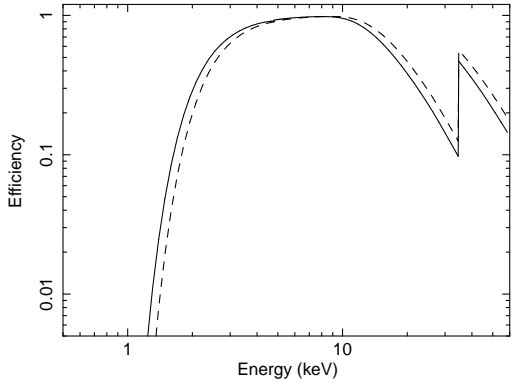

\begin{center}
\rotatebox{270}{
\FigureFile(50mm,){gsc_eff_v2.ps}
}
\end{center}
\caption{
X-ray detection efficiency of MAXI proportional counter for 
energy; a solid line indicates the efficiency for normal incident X-rays, while 
a dashed line indicates that of incident X-rays from 40 degrees.
}\label{fig:gsceff}
\end{figure}

 Considering this directional performance, we conducted various laboratory 
 tests of all proportional counters. Here we explain some of the results.
   The position resolution was tested at every 2 mm segment 
 of two-dimension on the incident window for all proportional counters.  
 The data were taken using 0.1 mm pencil X-ray beams with  X-ray 
 energies of 4.6, 8.0, and 17.4 keV.  The total length of the carbon anode 
 wire is 32 cm; thus, the total resistance is 31 to 37 
 k$\Omega$.   The  difference of resistance depends on the slight 
 difference of wire diameters.  The energy response also differs for 
 different anode wires.   We  took comprehensive data 
 useful for energy response and position response (Mihara et al. 
 2001; Isobe et al. 2004).  We also took the slat collimator 
 response data using X-ray pencil beams (Morii et al. 2006).  The 
 data-base, including all these data, is regarded as the response 
function at  the time of data analysis.
  In actual data analysis we will use each response function for the 
 two dimensional surface of all carbon wires. Pulse height response 
 depends on X-ray energy, where hard X-rays suffer the effect of 
 anomalous gas amplification (Mihara et al. 2001).

The simultaneous background for a localized source is measured 
in two regions separated from the source on the same detector 
in the FOV direction. 
Another background is measured in two regions
  just before and after observing the source  on the scanning   
path, where the two regions are  separated from the source.   
 In this case, both measurement locations of the  
background on the detector are the same as the  measurement  
location on the detector  of the source. 
The background is employed if  there are no appreciable  sources in 
these directions.   In addition, intrinsic  background is 
necessary to obtain the cosmic 
diffuse background.    The intrinsic background is gradually 
changing on orbit and for time even in the same 
direction (Hayashida et al. 1989).   
There is the small portion covered by a window frame of each detector 
in which cosmic diffuse X-rays as well as source X-rays are never 
irradiated.   A small portion of the North Pole direction is also
shadowed  by ISS structure.  Although these intrinsic background
count rates are not enough to measure for instant time, the data 
accumulated for days or more could be  used  to make a reasonable
 model of intrinsic background as obtained for Ginga proportional
counters (Hayashida et al. 1989).
\begin{figure}
\begin{center}
\rotatebox{270}{
\FigureFile(50mm,){expoprof.ps}
}
\end{center}
\caption{
A solid line shows the total product of effective area and dwelt 
time according to one orbit scan for a point source for the angle ($\beta$) 
slanted from  scanning a great 
circle.  This value is common to GSC-Z and GSC-H.   A chain line 
indicates the contribution of a central GSC unit only, while  
a dashed line and a dotted line are those of a right side GSC unit and 
a left side GSC unit only, respectively.
}\label{fig:gscarea}
\end{figure}
\subsection{Solid-state Slit Camera (SSC)}
\begin{figure}
\begin{center}
\FigureFile(85mm,){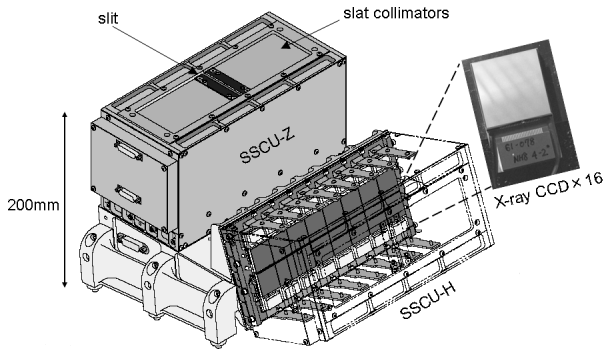}
\end{center}
\caption{
  SSC consisting of a horizontal SSC (SSC-H) 
 and a zenithal SSC (SSC-Z) and a CCD chip is depicted separately.
}\label{fig:ssc}
\end{figure}

\begin{figure}
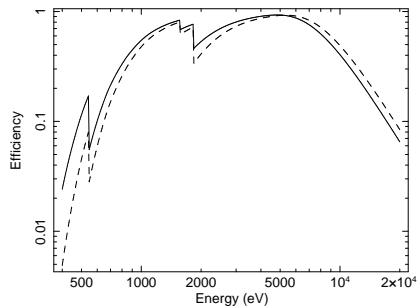

\begin{center}
\rotatebox{270}{
\FigureFile(40mm,){fig8.ps}
}
\end{center}
\caption{
 X-ray detection efficiency of MAXI CCD for energy; 
solid line is the efficiency for a normal incident X-ray, 
while a dashed line indicates that for  X-rays incident from 40 degrees.
}\label{fig:ssceff}
\end{figure}

  Each unit of SSC-H/Z consists of 16 CCDs, where each CCD acts as a 
 one dimensional position sensitive detector for the slat collimator 
 and the slit hole similar to the GSC system presented  in Figure 1 
and Table  1 (Tomida et al. 2009). The SSC-H camera is tilted up 16 
degrees so as not to view the upper atmosphere.  The response of the slat 
collimator scanning direction is 3 degrees 
 for bottom-to-bottom, while a slit image corresponds to 1.5 degrees 
 covering a wide FOV of 90 degrees as shown in Figure 7.  
 One orbit scan of the SSC corresponds to 90 degrees x 360 
 degrees.  Thus, it takes  SSC 70 days to scan the entire sky depending on the  
 precession of the ISS orbital plane, except for the bright region 
 around the Sun.  Thus, SSC will acquire actual all sky images  
 every half  year.

  The X-ray CCD chip of SSC is produced by Hamamatsu Photonics 
 K.K..  The CCD depletion layer is about 70 $\mu$m, of which the 
 X-ray detection efficiency is indicated in Figure 8.  The observation 
energy band of SSC is set to be 0.5-12 keV in standard mode, but 
observations above 12 keV are possible in a special mode.
However, since SSC is characterized by a low energy band,  
the energy band may be slightly shifted to lower energy than 0.5 
keV, depending on the temperature for thermal noise.  The CCD with
1024$\times$1024 pixels, and a pixel size of 24 $\mu$m $\times$ 24 $\mu$m is a 
two-dimensional
array, but SSC requires only one-dimensional position information.
Hence, multiple rows are summed in a serial register at the bottom of the imaging
region, and the summed charges in the serial register are transferred to 
a read-out node. Eight, 16, 32 or 64 summed rows  can  be selected in 
normal observation by commands. The larger this number, the better the time 
resolution, and the better the angular resolution in the X-ray sky map. 

 The CCD is sensitive to  particle radiations, 
 but an irradiation test of the simulated 
 radiation belt fluence suggests that the CCD can survive  
 for the expected three-year mission (Miyata et al. 2003).  
 Furthermore, it is possible to inject charges in the CCD before 
 it becomes damaged due to irradiation (Miyata et al. 2003).  
 Charge injection effectively  restores the degraded
 performance of X-ray CCDs (Tomida et al. 1997).

  To achieve better energy resolution with a CCD, all CCDs are 
 cooled  to -60 degrees C using thermo-electric coolers (Peltier 
 devices) and the LHPRS.  The maximum 
 power of the thermo-electric cooler is 1 W/CCD.  The LHPRS can 
automatically  transfer heat from the Peltier device and emit heat from 
the radiation panel.   Performance tests in the laboratory have been 
conducted with satisfaction 
 (Miyata et al. 2002; Katayama et al. 2005). The nominal energy resolution 
width of X-ray spectra is 150 eV for Mn K X-rays (5.9 keV).

\subsection{Support Sensor}

  Since the ISS has a huge structure of  $\sim$108 m 
and $\sim$420 tons, the 
 attitude determined in a certain part of the ISS may be slightly
 different from that in a distant part, due to shaking of the structure.  
MAXI is designed to 
 determine a target location as precisely as less than 0.1 degree. 
 Thus, it is necessary to determine the attitude of the MAXI coordinate 
 by less than 0.1 degrees every time.  For this purpose, MAXI itself 
 has a VSC and an RLG.  The VSC 
 can observe three or more stars, and it can determine a MAXI coordinate 
by 0.1 arc minute.  When the VSC is not 
 available for measurement due to solar radiation, the RLG will 
 extrapolate the attitude from a certain result of the VSC to the 
 following result.  This attitude determination is performed 
automatically by onboard software including a Kalman filter 
 on the MAXI data processor (Ueno et al. 2009; Horike et al. 2009).

  The high voltage of a proportional counter of GSC will be reduced 
 to  0 volt when solar radiation or SAA radiation is extremely 
 strong.  The command signal for this reduction will be issued by 
 working of the Radiation Belt Monitor (RBM).  This 
 high voltage reduction is also useful for setting program commands 
 in advance because of the possible prediction of solar position and 
 SAA location.

   The time precision of photon acquisition from GSC is 0.1 msec 
   by referring 
   to GPS signal. This precise absolute time is used for milli-second 
   pulsars and burst acquisition analysis.  Thus, only one GPS is 
   installed in MAXI for precise time reference without unknown lag.
   However, the time resolution of SSC is 5.8 sec in the normal
observation mode.

\section{MAXI Simulation and Expected Performance}

 A realistic simulation has been conducted with the recently  
developed  MAXI simulator, which generates fully simulated data 
of MAXI instruments 
on the ISS (Eguchi et al. 2009). The simulator takes into account 
various conditions on 
the ISS, i.e., the occultation of the sky with solar panels, 
particle and intrinsic background, and response function of X-ray 
cameras.  The attitude data and absolute time are attached to each 
event data on the ground. Thus, we can plot each event with energy 
information in a certain direction in the sky.  Integrated events 
from a certain direction for a certain time correspond to intensity, 
including background from the direction.  Here we demonstrate some of 
the simulated results (Hiroi et al. 2009; Sugizaki et al. 2009).
\begin{figure*}
\begin{center}
\FigureFile(50mm,){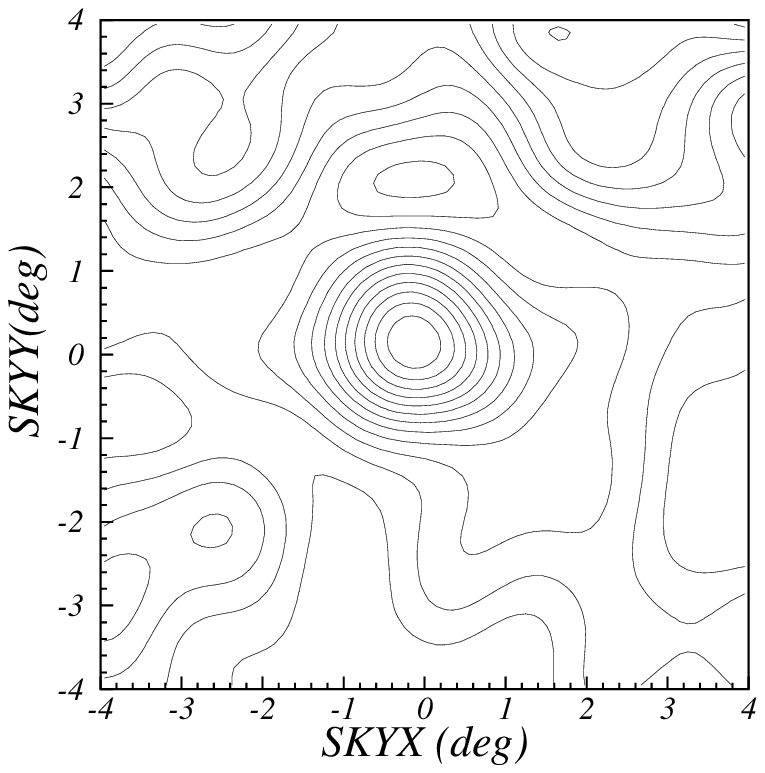}
\FigureFile(50mm,){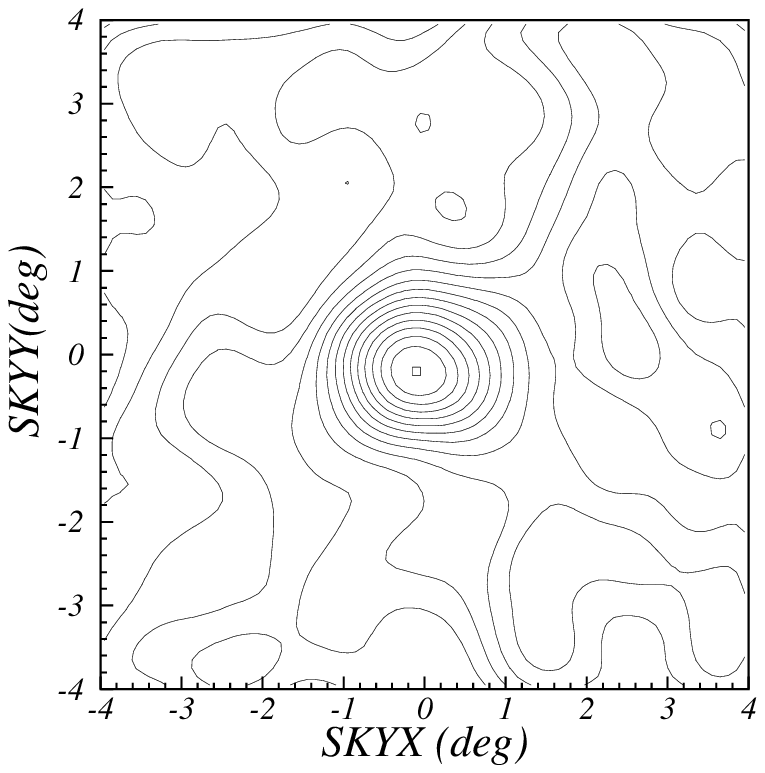}
\\
\FigureFile(50mm,){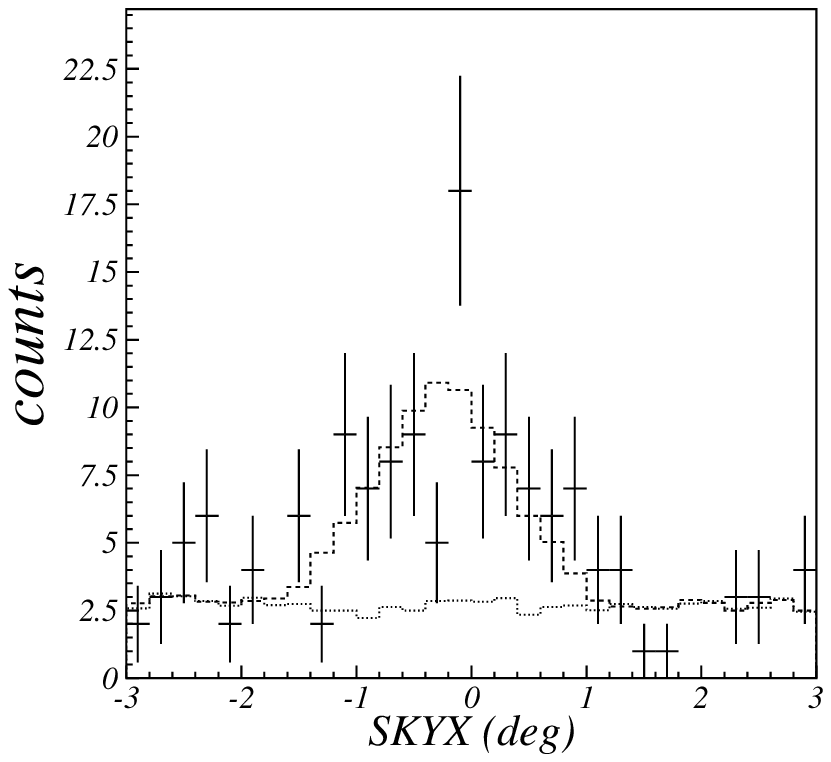}
\FigureFile(50mm,){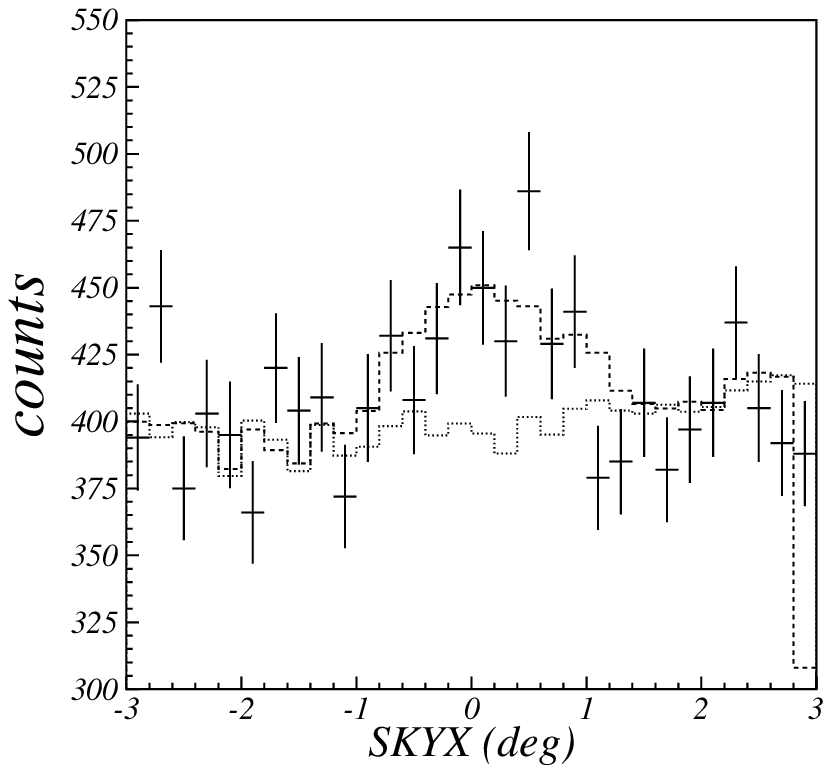}
\end{center}
\caption{ 
(a) Simulation is used to detect a source of 20 mCrab for  
one ISS orbit. 
Contours are smoothed in  the upper panel, while a cut figure through 
a peak is indicated with data points in the lower panel.
(b) Simulation is used to detect a source of 2 mCrab in one week.
Contours are smoothed in the upper panel, while a cut figure through
a peak is indicated with data point in the lower panel.
}\label{fig:src}
\end{figure*}
\begin{figure}
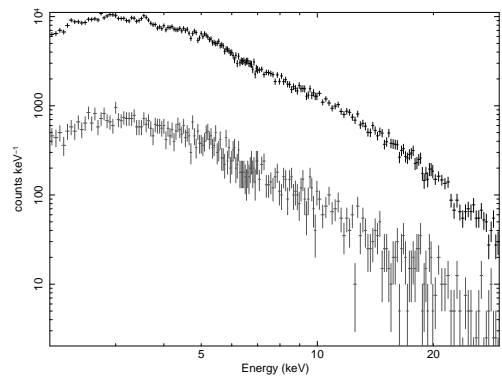

\begin{center}
\rotatebox{270}{
\FigureFile(50mm,){crab_pi_2.ps}
}
\end{center}
\caption{ 
Simulated spectra of Crab nebula for one-orbit (lower spectrum) and one-day 
(upper one) observations.
}\label{fig:spec}
\end{figure}

 First, we investigate the detectability of the GSC.  Figure 9 
illustrates one-orbit and one-week observations.  The results indicate 
that we can detect a  20 mCrab source in one-orbit at a 
confidence level of 5-sigma, and a 2 mCrab source in one-week at a 
 level of 5-sigma.  These simulations are estimated for the 
energy band of 2-30 keV of the Crab nebula spectrum.  This simulation is  ideal, because the
particle and intrinsic background 
of 10 counts sec$^{-1}$counter$^{-1}$ is 
steady, except for  statistical fluctuation during the 
observation time, and this background is subtracted with statistical 
fluctuation from integrated events of the source.  The spectrum of an 
intrinsic background is referred to the laboratory test result.  
In reality, we must consider the changeable background on the ISS orbit 
and systematic errors for 
attitude determination and response functions.  The final detectability 
could be 0.2 mCrab for a two-year observation, which means a source confusion limit for the angular resolution of 1.5 degrees.  
\begin{figure*}
\begin{center}
\FigureFile(130mm,){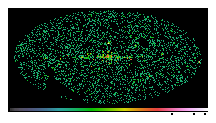}
\FigureFile(130mm,){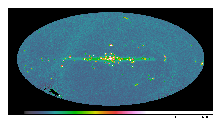}
\end{center}
\caption{ 
An all-sky X-ray source image (raw data) simulated for one-orbit   
observation (upper panel) and an image  (exposure corrected) for
one-month  (lower panel) on the Galactic coordinate.  The 
undetected (dark) region on the left bottom corresponds 
to the solar direction.  The two circular dark regions on the
upper panel correspond to unobservable regions around both end
directions of ISS pitch axis.
}\label{fig:allsky}
\end{figure*}

Here it is noted again that the above 5-sigma detectability is under 
the best condition for known source.  The detectability depends on
the intrinsic background or trapped particle background.
Either of zenithal and horizontal cameras can observe most directional 
sources even if the ISS passes through the SAA.   This makes
the detectability uniform for most direction except the direction 
around the Sun if observations are repeated by orbits.  In fact, 
a deviation from uniformity of the detectability of one orbit 
becomes worse by about 65 $\%$ 
for the sources observed on the SAA path, while that of
one day becomes uniform within 10-20 $\%$  for any direction 
except the solar region.  We evaluated another 
detectability excluding  observation in the region less than 
the cut-off rigidity of 8 (e.g., aurora region).  
A situation of this detectability 
is  not so much different from that of the aforementioned detectability.   Lastly, it is noted that a detectability for unknown   
source is not significantly different from  that for known sources.
One-sigma location error for unknown 
5-sigma sources is about 0.2 degrees for both one day and one  
week observations.

 Second, we perform a simulation to derive the energy spectra of a 
source of 1 Crab for a one-orbit and a one-day observation as shown 
in Figure 10.  This simulation is also ideal, because the 
background is steady only with statistical fluctuation and 
ideal determination of the attitude without an unknown systematic error
is assumed.  Nevertheless, it is concluded that reasonable spectra from a considerable number of sources are obtained by MAXI.

 Another important goal of MAXI is to make X-ray source 
catalogues for every period of  concern.  We can make light curves with the 
time bin for various X-ray sources.  The light curves of most 
Galactic X-ray sources are obtained with time bins of one orbit or 
one day.  If the time bin is one month, it is possible to make 
light curves of a considerable number of AGNs brighter than 1 mCrab 
as shown on the lower panel in Figure 11.  It is not easy at a  
 glance to discriminate 
between  the weak sources and the CXB, but 1-2 mCrab sources are  
significantly detected as simulated in Figure 9 (b).   If we integrate 
all event data from the sky 
for two years, we can create an X-ray source catalogue brighter 
than 0.2 mCrab at a confidence level of 5-sigma.  
In general, since the  time scale of variability of AGNs is longer than 
that of Galactic sources, catalogues with time bins of one 
month and one year will provide new information about long-term variable 
sources such as the super massive Binary Black Hole (BBH).  Actually, 
an all-sky 
image for every six-month observation is compatible with the result
 of  HEAO-1 A2.    

\section{Concluding Remarks}

 MAXI is the first payload to be attached to JEM-EF (Kibo-EF) 
of the ISS. It consists of two kinds of X-ray cameras: the Gas Slit 
Camera (GSC), which has one-dimensional position sensitive 
gas-proportional counters with a 2 to 30 keV  X-rays 
energy band, and the Solid-state Slit Camera (SSC), which has X-ray CCDs 
with a 0.5 to 12 keV X-rays energy band. 
MAXI provides an all-sky X-ray image acquisition 
capability under the best conditions for 
every ISS orbit. If MAXI scans the sky for one week, it could make 
2 mCrab X-ray all sky image excluding the bright region around the 
Sun.  Thus, MAXI  not only rapidly informs astronomers worldwide 
about X-ray novae and transients  if  they occur, but also observes 
long-term variability of Galactic and extra-Galactic X-ray 
sources.  At the same time, MAXI  provides an X-ray source catalogues
accumulated for various periods such as six months, one and more years.

All MAXI instruments  are ready for launch. Data 
processing and analysis software, including an alert system on the 
ground, are being developed by the mission team.  MAXI will be attached to 
JEM(Kibo)-EF in  mid-2009.  Anyone can participate in follow-up 
or multi-wavelength observations of MAXI objects with ground 
optical and radio telescopes, and at space observatories.  Please 
contact the present authors  for more information and further collaboration.

Finally, please refer to the Proceedings of Astrophysics 
with All-Sky X-Ray Observations -- 3rd 
International MAXI Workshop --, RIKEN, Wako, Japan, June 10-12, 2008, 
ed., N.Kawai, T.Mihara, M.Kohama and M.Suzuki 
(http://cosmic.riken.jp/maxi/astrows/) 
for further information on MAXI instrumentations and preliminary guideline 
for MAXI data.
\\

 MAXI fabrication and testing are also supported by the staffs at the 
JEM project office and the Space utilization office.  We thank 
them for their valuable guidance and cooperation in maximizing the scientific
output from MAXI.  We also thank the former colleagues who developed and
supported this long-term project of MAXI, although these many colleagues 
(staffs and graduate students) have not been included as  co-authors.

 We also thank the following companies and institutions for developing 
MAXI sub-systems:  NEC Co. for the MAXI bus 
system, Meisei Electric Co. for the mission instruments, Oxford 
Instrum. Co. and Japan I.T.S. Co. for the gas proportional counters, 
Hamamatsu Photonics K.K. for the CCDs, ATK Space (former Co.:  
Swales Aerospace) for 
the LHPRS, DTU (Technical University of Denmark)  for the Optical Star 
Sensor, the Institute of Aerospace Technology in JAXA for the GPS, and  
other cooperating companies and institutions.

\end{document}